\documentclass[10pt, conference, compsocconf]{IEEEtran}

\usepackage{graphicx}
\usepackage[caption=false]{subfig}
\usepackage{cite}

\let\labelindent\relax

\usepackage{enumitem}

\usepackage{amsmath}
\usepackage{algorithm}
\usepackage{algorithmicx}
\usepackage[noend]{algpseudocode}
\usepackage{amsthm}
\theoremstyle{definition}
\newtheorem{definition}{Definition}
\begin{document}
\bstctlcite{IEEEexample:BSTcontrol}
%

\title{Proactive Composition of Mobile IoT Energy Services}


\author{\IEEEauthorblockN{Abdallah Lakhdari}
\IEEEauthorblockA{School of Computer Science\\
The University of Sydney, Australia\\
abdallah.lakhdari@sydney.edu.au}
\and
\IEEEauthorblockN{Athman Bouguettaya}
\IEEEauthorblockA{School of Computer Science\\
The University of Sydney, Australia\\
athman.bouguettaya@sydney.edu.au}
} 

\maketitle


\begin{abstract}
We propose a novel \textit{proactive} composition framework of \textit{wireless} energy services in a crowdsourced IoT environment. We define a new model for energy services and requests that includes providers' and consumers' mobility patterns and energy usage behavior. The proposed composition approach leverages the mobility and energy usage behavior to generate energy services and requests proactively. Preliminary experimental results demonstrate the effectiveness of generating proactive energy requests and composing proactive services.

\end{abstract}

%
\IEEEpeerreviewmaketitle

\section{Introduction}
The proliferation of the Internet of things (IoT), particularly wearables, may give rise to a self-sustained crowdsourced IoT ecosystem \cite{atzori2010internet}. The augmented capabilities of IoT devices such as sensing and computing resources may be leveraged for peer-to-peer sharing. People can exchange a wide range of IoT services such as computing offloading, hotspot proxies, {\em energy sharing}, etc. These crowdsourced IoT services present a convenient, cost-effective, and sometimes the only possible solution for a resource-constrained device \cite{ahabak2015femto}. For instance, a passenger's smartphone with low battery power may elect to receive energy from nearby wearables {\em using Wifi} \cite{raptis2019online}. The focus of this paper is on crowdsourcing IoT energy services. 


The concept of {\em wireless crowdsharing} has been recently introduced to provide IoT users with power access, anywhere anytime, through crowdsourcing\cite{raptis2019online}\cite{Previouswork11}. We leverage the service paradigm to unlock the full potential of IoT energy crowdsourcing. We define {\em an IoT Energy Service} as the abstraction of energy wireless delivery from an IoT device (i.e., {\em provider}) to another device (i.e., {\em consumer}) \cite{lakhdari2020composing}. Crowdsourcing IoT energy services has the potential of creating a \textit{green} service exchange environment by \textit{recycling} the unused IoT energy or relying on \textit{renewable} energy sources. For example, an IoT device may share its spare energy with another IoT device in its vicinity. Another example, a smart shoe may harvest energy from the physical activity of its wearer~\cite{choi2017wearable}\cite{gorlatova2014movers}. Additionally, wireless charging allows energy crowdsharing to be a \textit{convenient} alternative as the devices do not need to be tethered to a power point, nor use power banks.  Crowdsourcing energy services can be deployed through already existing wireless power transfer technologies such as \textit{Energous}\footnote{https://www.energous.com/} which can deliver up to 3 Watts power within a 5-meter distance to multiple receivers. 


{\em Service composition} is expected to play a vital role in the crowdsourced IoT environment. A single IoT energy service may not fulfill the requirement of a consumer due to the limited resources of wearables \cite{lakhdari2020Vision}. Instead, multiple services need to be composed to provision one energy request. Existing energy service composition frameworks mainly consist of the real-time discovery and selection of nearby energy services to accommodate an energy request. These techniques assume that a request can be fulfilled by the available energy services within its vicinity \cite{lakhdari2020fluid}. The focus of these composition techniques was only on the spatio-temporal composability \cite{lakhdari2020composing} and addressing the challenges related to the energy fluctuation, the mobility, and the conflicts of the available services \cite{lakhdari2020Elastic}\cite{abusafia2020reliability}\cite{chaki2020conflict}. 

\begin{figure}
\centering
\includegraphics[width=0.6\linewidth]{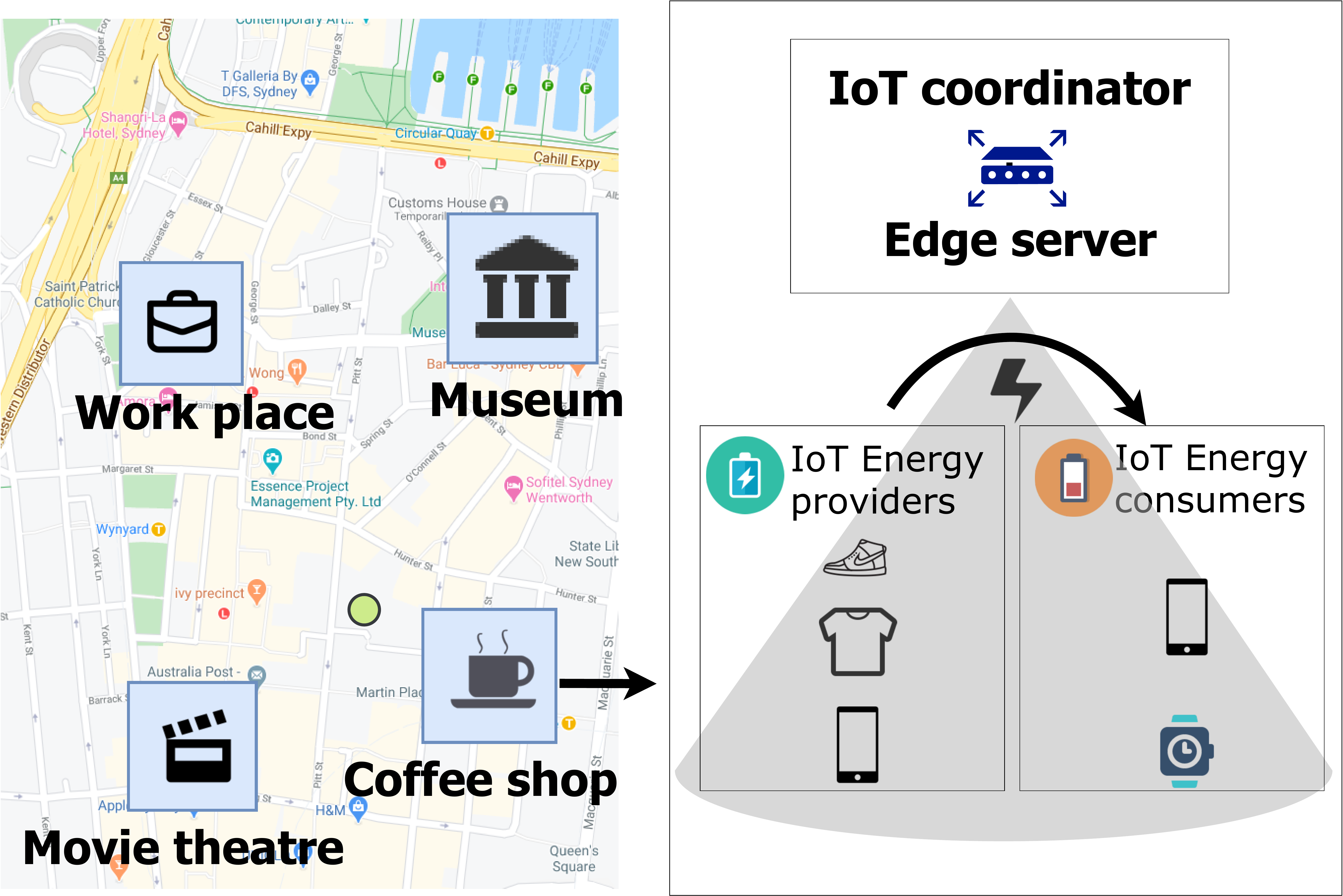}

\caption{\small Crowdsourcing IoT energy services}
\label{fig:MicrocellSharing}
\end{figure}

The crowdsourced IoT energy ecosystem is a \textit{dynamic} environment that consists of providers and consumers congregating and moving across \textit{microcells} boundaries. A microcell is any confined area in a smart city where people may gather (e.g., coffee shops, restaurants, museums, library), see figure \ref{fig:MicrocellSharing}. Energy service providers and consumers may have different spatio-temporal preferences. These preferences' incongruity may severely impact the \textit{balance} between energy services and existing energy requests. For instance, in a particular location at a specific time, an energy consumer might not find the required energy amount to fulfill their request. We propose a novel model for \textit{mobile crowdsourced energy services}, a major extension of the deterministic energy service model in \cite{lakhdari2020composing}. According to extensive experiments, it has been proven that human mobility is highly predictable \cite{gonzalez2008understanding}. This predictability can be related to the rough regularity of our daily routines  \cite{yang2015mobility}. We leverage these mobility patterns and the energy usage behavior of energy consumers and providers to propose a proactive composition approach. The proposed approach (i) proactively defines services and requests for energy providers and consumers, respectively, according to their spatio-temporal preferences and energy requirements. (ii) For each consumer, the proactive composition approach defines an optimal set of energy requests ahead of time after estimating their spatio-temporal preferences, their energy usage behavior, and the energy availability likelihood at each microcell.


We propose a \textit{proactive} service composition framework that captures and leverages the spatio-temporal preferences of energy providers and consumers. The proactive service composition identifies, for an energy request, \textit{when}, \textit{where}, and \textit{which} set of energy services to compose. \textit{The goal is to avoid the under-provision of energy requests and balance the energy requests and services within and across microcells}. Our proposed solution's novelty is that energy services are composed seamlessly without affecting the usage behavior and the mobility patterns of service providers and consumers. First, the proposed composition framework anticipates the next required energy based on the energy usage behavior. The proactive composition, then, plans, ahead of time, the location, time, and required amount of one or multiple energy requests according to the consumer's mobility pattern. Additionally, the proposed composition approach considers the availability of energy services within the microcells to be passed by the consumer. The composition framework determines the availability of energy services based on the energy usage behavior and the mobility patterns of the providers. The preferred solution is {\em to select an optimal set of services and fulfill the requirement of an energy consumer without affecting the spatio-temporal preferences of the providers and consumers}. The main contributions are:  
\begin{itemize}[ noitemsep,nosep,leftmargin=15pt,labelsep=5pt,itemindent=0pt, labelwidth=*]
    \item A novel model to represent the proactive energy services and requests by leveraging IoT users' spatio-temporal preferences and their energy usage behavior.
    \item A proactive composition framework  that automatically matches proactive service with  proactive energy requests in a crowdsourced IoT environment.
  
\end{itemize}

\section{Motivating scenario}
\begin{figure*}
\centering

\includegraphics[width=\textwidth]{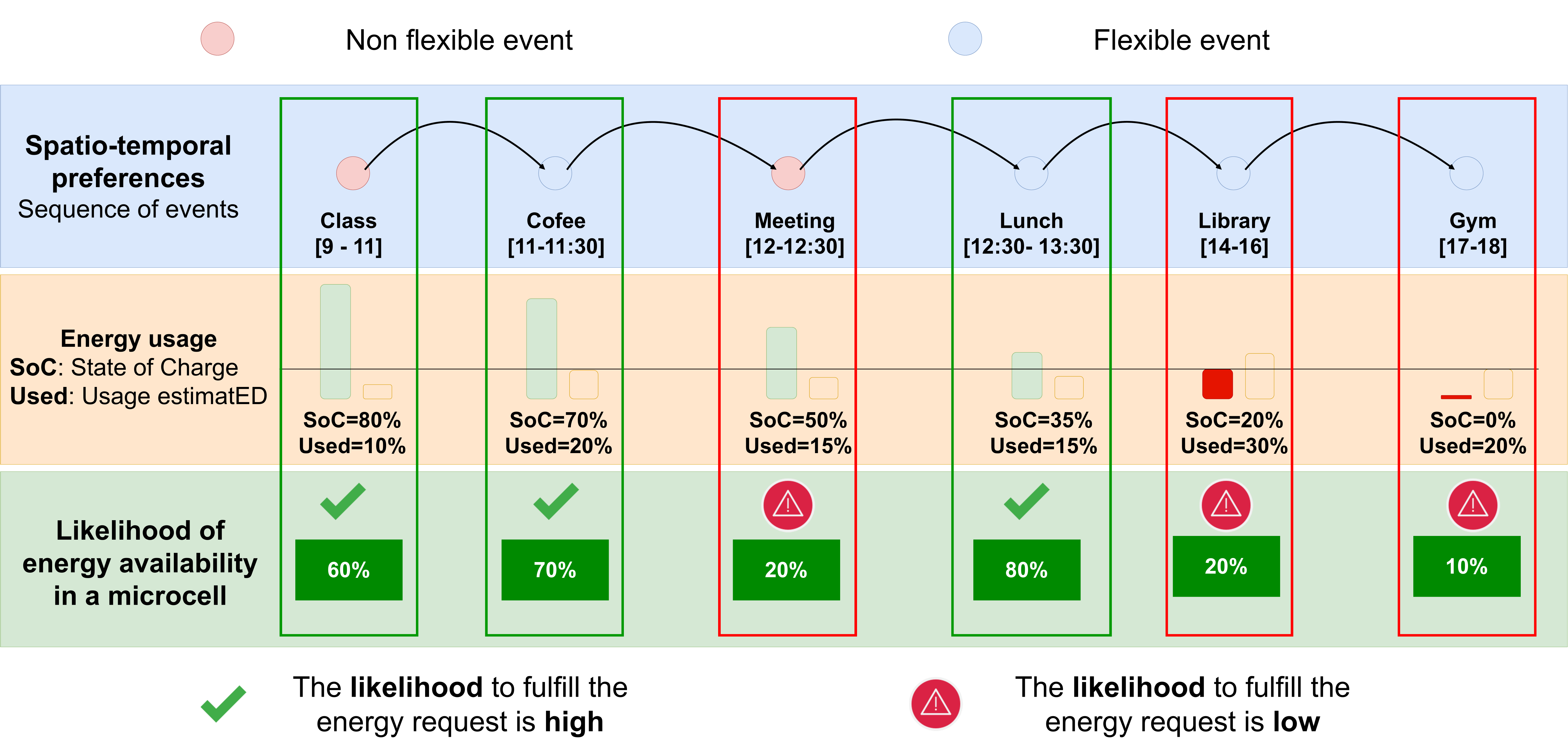}

\caption{\small Student's daily routine within the campus, spatio-temporal patterns and energy usage behavior}
\label{fig:MTVSCN}
\end{figure*}

We describe a scenario at the university campus where students move across, gather, and practice various activities in different confined places (i.e., microcells), e.g., library, study spaces, food court. We assume that students are incentivized to recycle the spare energy of their IoT devices by sharing it with nearby devices \cite{abusafia2020incentive}. Energy providers advertise their energy services in different microcells within the smart campus according to their spatio-temporal preferences and their energy usage behavior. The advertisement presents a description of the provided service, e.g., location, start and end time, the provided energy amount. A centralized IoT coordinator processes all energy services, i.e., an edge server responsible for sharing crowdsourced IoT energy services (see figure \ref{fig:MicrocellSharing}). Typically, in an energy crowdsharing environment, whenever a student needs to recharge their smartphone, they launch an energy request in a microcell in the following form: \textit{User $x$ requires an amount of energy $E$  in the location $L$ during the period $[st~,~et]$}. The IoT coordinator then composes the available nearby energy services to fulfill the said energy request \cite{lakhdari2020composing}. However, sometimes, it is unlikely to find the required energy amount in a specific location or a specific time.

Figure \ref{fig:MTVSCN} presents a student's daily routine within the campus coupled with their energy usage behavior. For example, the student has a class from 9 to 11 am, a 30 minutes coffee break after the class, then an important meeting with their supervisor, lunch break, and finally a 2 hours study session in the library before heading to the gym at the end of the day. The daily routine is drawn as a sequence of spatio-temporal events. Each event is represented by location, time frame, and the student's smartphone energy status (i.e., battery state of charge and the estimated required energy). The student's smartphone is set to automatically launch an energy request to nearby IoT devices whenever the battery state of charge reaches a predefined threshold, e.g., 20\%. In this case, an energy request would be launched by the student's smartphone in the library at around 4 pm to crowdsource nearby energy services \cite{lakhdari2020composing}. Unfortunately, it is unlikely for the student to fulfill their energy requirement at this time in the library due to a supply shortage. This shortage can be related to several reasons; for instance, energy providers may run out of their spare energy. Another reason could be that multiple devices are requesting energy to ensure the functioning of their devices for the rest of the day.

A better solution to crowdsource energy is to leverage the mobility patterns of students with their energy usage behavior to \textit{proactively} define \textit{when}, \textit{where}, and \textit{how much energy} to request to fulfill their energy requirement. Leveraging the student's spatio-temporal preferences and their flexibility would increase the likelihood of providing the requested energy along with an insignificant alteration of the student's daily routine. Instead of launching an energy request re-actively whenever a State of Charge (SoC) threshold is reached \cite{Previouswork11}, the proactive composition selects the optimal time, location, and energy amount to request to fulfill their requirement. For instance, it can be easily anticipated that the smartphone's battery at 4 pm in the library would not suffice their energy requirement based on the student's smartphone's estimated energy consumption behavior (see figure \ref{fig:MTVSCN}). Figure \ref{fig:MTVSCN} also depicts the energy availability likelihood at each microcell visited by the student requesting energy. In that case, the energy provision likelihood in the library is estimated to be very low (e.g., 20\%) according to the energy usage and provision behavior of the existing providers in the library at that time. However, it is shown that the available energy in other microcells visited earlier by the student according to their daily routine, namely, 70\% in a visited coffee shop and 80\% in the food court at lunchtime.  The framework leverages the knowledge about the energy availability across microcells, the energy consumption behavior, and the spatio-temporal preferences of the student requesting energy to define one or multiple energy requests in the class, the coffee shop, and the food court, respectively, to fulfill the anticipated energy requirement.

\section{Proactive energy services and requests}
An energy service is the abstraction of the wireless delivery of energy from an IoT device to another. We extend the existing energy service model to a proactive energy service model \cite{lakhdari2020composing}. The proactive service model leverages the spatio-temporal patterns of energy providers along with the usage behavior of their IoT devices to seamlessly define, advertise, and share their spare energy without affecting their daily routine preferences. Each individual may concur with a pattern for their spatio-temporal preferences  \cite{gonzalez2008understanding}. They might regularly visit particular places at specific times. These habits may define the {\em mobility patterns} of individuals \cite{yang2015mobility}. We leverage mobility and energy usage patterns to define a model for proactive energy services and requests.


\subsection{Proactive energy service model}
We define an energy provider as an IoT user who is willing to share the spare energy of their devices with nearby IoT devices. We use \textit{energy services} to abstract the energy sharing process. An energy service is the abstraction of the wireless delivery of energy from an IoT device to another \cite{Previouswork11}. In the previous energy service model. Energy provision was under the complete control of the IoT device owners. They \textit{manually} advertise their services \cite{lakhdari2020composing}. 

We extend the existing energy service model by formally representing the spatio-temporal patterns and the energy usage behavior of providers. The goal is to \textit{automatically} define the optimal time, location, and energy amount to provide according to the daily routine of the provider. For example, if a provider $i$ usually spends half an hour in a particular restaurant $L_i$ at lunchtime $T_i$ and their battery state of charge is $SoC_i$, the proposed model would capture the spatio-temporal pattern $<L_i, T_i, SoC_i>$. In addition to capturing this pattern, the proactive energy service model estimates the energy usage distribution and verifies if it is enough energy for the rest of the day. Accordingly, the service model defines the energy amount to share $E_i$ while staying at that restaurant.
We formally define a proactive energy service as follows:
\begin{definition}{\textbf{A proactive energy service}} \emph{ $ PES $ is represented as a tuple $< S, P, F , Q, M, U>$, where: }
\end{definition}
\begin{itemize}
    \item $S$ is a unique ID for the energy service 
    \item $P$ is a unique ID for the provider\footnote{In the rest of the paper, we assume that an energy provider shares energy only from one IoT device. Hence, Energy providers refers to both the device sharing energy and the owner of the device}.
    \item $F$ is the set of $PES$ functionalities offered by the service,
    \item $Q$ is a tuple of $<q_1, q_2, ..., q_n>$, each $q_i$ denotes a QoS property of $PES$. \cite{Previouswork11}. 
    \item $M$ is the mobility patterns of the provider $P$.
    \item $U$ represents the energy usage behavior of a provider $P$. 
\end{itemize}

\begin{definition}{\textbf{The quality attributes of an energy service}} 
\emph {QoS is a set of parameters that allows energy consumers to distinguish among IoT energy services \cite{lakhdari2020composing}, namely:}
\end{definition}
\begin{itemize}
    \item $l$ is the location of the consumer. 
    \item$r$ is the range for a successful wireless energy transfer. 
    \item $st$ and $et$ represent the start time and end time of an energy service respectively. 
    \item $DEC$ is the deliverable energy capacity. 
    \item $I$ is the intensity of the wirelessly transferred current. \item $Tsr$ represents the transmission success rate. 
\item $Rel_i$ is the reliability of the provider.  
\end{itemize}


\subsection{Proactive energy request model}
An energy consumer in the crowdsourced IoT environment is an IoT user who requests energy for their devices from nearby IoT devices. We define an \textit{energy request} as an abstraction of the energy consumer requirement according to their spatio-temporal preferences \cite{Previouswork11}. In the existing energy service model. Energy consumption is under the complete control of the IoT device owners. They \textit{manually} request energy according to their preferences \cite{lakhdari2020composing}. 

Defining energy requests automatically, based on the anticipated mobility and usage behavior, increases the chances to fulfill the requirement of a consumer. For instance, a consumer might reserve proactive energy services ahead of time according to their preferences and the availability of services within the microcells visited by the consumer.

\begin{definition}{\textbf{A proactive energy service request}}
\emph{is a tuple $Rq< t, l, RE, CI, du>$, where:} 
\end{definition}
\begin{itemize}
    \item $ t$ is the timestamp when the energy request is launched. 
    \item $ l $ is the location of the consumer. We assume that the consumer is staying at their location $l$ within one microcell after launching their energy request. 
    \item $RE$ represents the required amount of energy. 
    \item $CI$ is the maximum intensity of the wireless current that a consuming IoT device can receive. 
    \item $du$ refers to the charging period \cite{Previouswork11}.
\end{itemize}

\subsection{Patterns of energy providers and consumers}
It has been proven that human mobility is highly predictable according to an extensive set of experiments on capturing the movement of millions of humans in metropolitan areas  \cite{gonzalez2008understanding}. The regularity in human mobility could be reflected on the IoT devices associated with their owners to define the mobility patterns of energy services users in a crowdsourced IoT environment \cite{lakhdari2020fluid}. Additionally, the energy consumption behavior of IoT devices reflects the usage behavior of the devices by their owners. A recent study that adopted a crowdsourcing-based method to model the energy consumption of smartphones revealed a certain regularity in the energy consumption behavior tightly correlated with the daily routine activities of the smartphone owners \cite{peltonen2015energy}. Similarly, we leverage the captured regularity in the energy consumption behavior to automatically define energy services and requests in the crowdsourced IoT environment. In what follows, we define the mobility and the energy usage patterns of energy providers and consumers.
\begin{definition}{\textbf{Mobility patterns }} 
\emph {The mobility pattern $M$ of an IoT user in the crowdsourced IoT energy environment is the captured spatio-temporal sequence of the IoT user activities during their daily routine.}
\end{definition}

The daily routine spatio-temporal sequence can be represented as a probabilistic time series that denotes at each timestamp the availability likelihood of the IoT user $i$ in a specific location $loc_{i}$ at a particular time $t_i$. Formally, the  mobility $M_i$ of an IoT user $i$ is a tuple $M_i(t_{i}, loc_{i}, \theta_{i})$.
\begin{itemize}
    \item $t_i=\{t_{i0},t_{i1},t_{i2},~..,t_{in}\}$ is the set of timestamps. In this work, we consider a five minutes time interval between two consecutive timestamps. 
    \item $loc_i=\{loc_{i0},loc_{i1},~..,loc_{in}\}$ is the set of visited locations by the IoT user $i$. 
    \item $\theta_{i}$ is the likelihood that the provider $i$ is at location $loc_i$ at time $t_i$.
\end{itemize}

Each $\theta_{ik} \in \theta_i=\{\theta_{i0},\theta_{i1},\theta_{i2},~..,\theta_{in}\}$ is the probability that IoT user $i$ is in the location $loc_{ik}$ at timestamp $t_ik$. $$\theta_i<l_i=loc_{i}>= f(H_i)$$

The IoT user's availability probability distribution function $f_i$ can be obtained by statistical methods applied on the historical records $H_i$ of the IoT user $i$ \cite{deng2016constraints}.

\begin{definition}{\textbf{Energy usage behavior}} 
\emph {The usage behavior $U$ of an IoT user is the captured state of charge of the user's battery during the day.}
\end{definition}

The IoT user's battery state of charge patterns are represented as a time series that denotes at each timestamp the estimated battery level 
according to their historical record of energy usage. Similarly, the energy usage behavior $U_i$ can be formally represented as a tuple $U_i(t_{i}, SoC_{i})$.
\begin{itemize}
    \item $t_i=\{t_{i0},t_{i1},t_{i2},~..,t_{in}\}$ is the set of timestamps. In this work, we also consider a five minutes time interval between two consecutive timestamps to capture the energy usage behavior. 
    \item $SoC_i=\{SoC_{i0},SoC_{i1},SoC_{i2},~..,SoC_{in}\}$ is the state of charge temporal sequence of the IoT users $i$. 
\end{itemize} 

\begin{definition}{\textbf{The flexibility of energy consumers}} 
\emph {is defined by their toleration to stick or change the anticipated microcell to visit at a specific time and their toleration to extend or shorten their stay time in that microcell}
\end{definition}

Typically, according to the activities of our daily routine activities, some of the visited places are unchangeable. For example, a student cannot change the location or the time of their class from 9 to 11 am. However, it is obviously possible to change the venue to have a coffee or a lunch (see red and blue circles in figure \ref{fig:MTVSCN}). Additionally, the interval of time spent in regularly visited places e.g., coffee shops or food courts can be defined as a {\em range} between the { \em minimum} and the {\em maximum} duration a person would stay \cite{do2013places}. We incorporate the flexibility of energy consumers based on their toleration to change the microcell and the range of their usual stay in that microcell.

The flexibility $Fx$ of  an  energy consumer $i$ is associated with their mobility pattern. It can be  modeled as a tuple $Fx_i<A_i, Ch_i, Sr_i>$  where:
\begin{itemize}
    \item $A_i$ is the availability distribution for all visited microcells $MCL$ during the daily routine of the energy consumer $i$.
    $$MCL_{ik} \in MCL_i=\{MCL_{i0},MCL_{i1},~..,MCL_{in}\}$$
     \item $Ch_i$ is the corresponding distribution that indicates for each microcell if it is tolerated to be changed.
    \item $Sr_i$ is the tolerance distribution that indicates  if the stay time can be extended or shortened in a visited microcell.
\end{itemize}
The availability distribution of a consumer $i$ can be derived from the mobility pattern of that consumer $M_i$. The toleration distributions $Ch_i$ and $Sr_i$ can be presented as a binary sequence of the  all the visited microcells $MCL$ during the daily routine of the energy consumer $i$ $MCL_{ik} \in MCL_i=\{MCL_{i0},MCL_{i1},MCL_{i2},~..,MCL_{in}\} $ as follows:   
$$Ch_{ik}=1 ~if~  MCL_{ik}~ is ~changeable;~~0 ~ otherwise $$
$$Sr_{ik}=1 ~if~  T <MCL_{ik}> ~is ~flexible;~~0 ~ otherwise $$
Where $T <MCL_{ik}>$ is the stay time interval of the energy consumer $i$ at the microcell $k$.

\begin{algorithm}[t!]
\footnotesize
    \renewcommand{\algorithmicrequire}{\textbf{Input:}}
    \renewcommand{\algorithmicensure}{\textbf{Output:}}
    \caption{Proactive composition algorithm}
    
    \label{alg:STcompo}
    \begin{algorithmic}[1]
        \Require
        $<$ $ Provider$, $Consumer$, $Microcell$ $>$ 
        \Ensure 
        $<Consumer plan>, ~<ER>$
        \State  \text{Define mobility plan for each provider $ProvPlan$}
        \State  \text{Define proactive services $<PES>$}
        \State  \text{Generate the mobility graph MG}
        \State  \text{Estimate the energy availability}
        \State  \text{Define proactive requests $<ER>$}
        \State  \text{Define mobility plan for each consumer $ ConsPlan $}
        \State \Return $ ConsPlan $
    \end{algorithmic}
\end{algorithm}

\subsection{Proactive composition framework}
The proposed framework utilizes the previous definitions in the composition as follows (See Algorithm \ref{alg:STcompo}): 
\begin{enumerate}
    \item According to the daily behavior of the energy provider, one or multiple energy services can be generated. The spatio-temporal quality attributes of these energy services (i.e., $l, st $ and $et$) are defined based on the mobility pattern $M$ of that provider, e.g., time spent in regularly visited places based on their daily activity model \cite{gonzalez2008understanding}. $DEC$ and $Rel_i$ are estimated by the energy usage model $U$ of their IoT device.  $I$ and $ Tsr$ are defined based on the specifications of the IoT devices sharing energy \cite{lakhdari2020fluid}. 
    
    \item Similarly, according to the daily routine of the energy consumer, one or multiple energy requests can be generated. The spatio-temporal features of these energy requests (i.e., $l, st $ and $du$) are defined based on the mobility patterns of the energy consumer. $RE$ is estimated by the energy usage model of the IoT device.  $CI$ is defined based on the specifications of the IoT devices requesting energy \cite{lakhdari2020composing}.
    
    \item The framework creates a mobility graph where the nodes  and the edges represent the microcells and the movement of IoT users between these microcells, respectively. 

    \item The framework estimates the energy availability at each microcell based on the generated proactive services and the estimated energy to be shared by each service. 

    \item The energy consumer then selects the microcells to visit according to the activity sequence of their daily routine and their anticipated energy requirement as follows: 

\begin{enumerate}
        \item The consumer first identifies the flexible and the non-flexible activities based on their mobility pattern \cite{lakhdari2020Elastic}. 

        \item The consumer identifies their energy requirements during the day based on their energy usage behavior.

        \item  The energy consumer defines their mobility plan in the form of a path within the energy availability graph.

        \item For each proactive energy request, the framework performs a spatio-temporal composition of the available energy services \cite{lakhdari2020fluid}.
    \end{enumerate}
\end{enumerate}

\section{Preliminary experiments}
We conduct a set of experiments to evaluate the proposed concepts of proactive service and requests and the proactive composition. We  assess the \textit{effectiveness} of the proactive composition approach by measuring the ratio of successful requests in and across different microcells.  We compare the proposed approach with a state-of-the-art  algorithm, i.e., the spatio-temporal composition of energy services \cite{lakhdari2020composing} and with a brute-force-based composition approach.
\begin{figure}
\centering
\includegraphics[width=\linewidth]{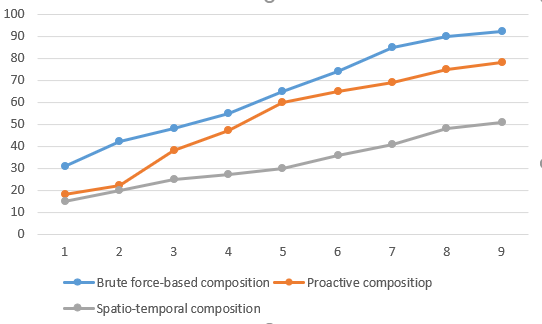}
\caption{\small Provisioning variation according to the energy availability}
\label{fig:res2}
\end{figure}

\subsection{Datasets and experiment environment}
%
\textit{To the best of our knowledge, it is challenging to find a dataset about the energy wireless transfer among human-centric IoT devices}. We create a crowdsourced IoT environment scenario close to reality. We mimic the crowd's energy harvesting and sharing behavior by utilizing \textit{QLD smartgrid}\footnote{https://data.gov.au/dataset} an energy sharing smart-grid of 25 houses in  Australia. In our experiment, we define the energy service QoS parameters, the deliverable energy capacity $DEC$, and the intensity of the transferred current $I$ from the \textit{QLD smartgrid} dataset. The transmission success rate $Tsr$ QoS parameter is randomly generated. Similarly, the energy requirement of a request $Rq.RE$ is also generated from the daily energy consumption of the houses.


We use {\em Yelp}\footnote{https://www.yelp.com/dataset} dataset to simulate the spatio-temporal features of energy services and requests. The dataset contains information about the crowd's behavior in different venues in multiple cities. Our experiment only focuses on people's check-ins information in confined areas, e.g., coffee shops. For each venue ($business\_id$), we extract the crowd size ($Checkins$) at each hour ($hour$) of the day ($weekday$). We assume these people as IoT users. They may offer energy services from their IoT devices while staying in a confined area. We define spatio-temporal features of energy services and requests by generating the check-in and check-out timestamps of customers to microcells. For example, the start time $st$ of an energy service from an IoT user is the time of their check-ins into a coffee shop.


\begin{figure}
\centering
\includegraphics[width=\linewidth]{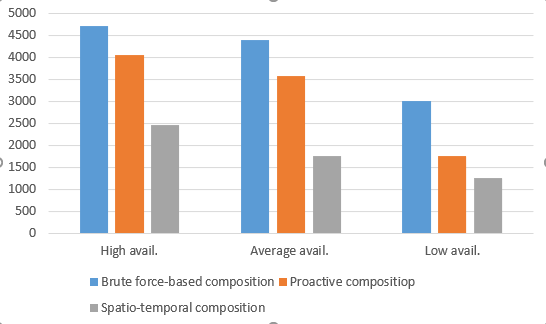}

\caption{\small Successful requests with different techniques}
\label{fig:res1}
\end{figure}
\subsection{Effectiveness} 
We investigate the effectiveness of the proactive composition by comparing the number of successfully served requests by the three composition algorithms (see figure \ref{fig:res1}). This initial comparison aims to mention the overall behavior of the three composition techniques in three different settings. Namely, high availability (i.e., six energy services for one request), average availability of energy across microcells (i.e., three energy services for one request), scarce energy availability  (i.e., one energy service for one energy request). As a baseline to compare with, the brute force composition technique requests energy from every visited microcell. It is expected to see the best behavior with this composition technique. The poor performance of the spatio-temporal composition is expected because the algorithm has not been designed to adjust according to the external environment. It is worth mentioning that with the change in the energy availability across microcells, the difference between the three techniques remains almost similar. The closeness between the brute force-based composition and the proactive composition is a positive indicator of the effectiveness of the proactive composition.


We run another experiment to investigate the effectiveness of adapting energy requests to the energy availability within and across microcells. We change the ratio of \textit{number of services/number of requests} from  1 to 9  and measure the received energy amount per request. Figure \ref{fig:res2} presents the behavior of the three composition techniques. There is an apparent closeness between the behavior of the brute force-based composition and the proactive composition with a remarkable difference with the performance of the spatio-temporal composition. However, the experiment shows no significant difference between the spatio-temporal and the proactive composition algorithms when the energy availability is low, namely when the ratio number of services/number of requests is between 1 and 2.5. After these values, the proactive composition behavior improves significantly. Adjusting the energy requirements and proactively redefining energy requests according to the availability of energy services and spatio-temporal preferences would increase the chances of fulfilling the consumer's requirements during their daily routine.

\section{Conclusion}
We propose a novel proactive spatio-temporal composition framework to crowdsource energy services from IoT devices. The goal is to seamlessly meet users' energy requirements in the dynamic crowdsourced IoT environment. We define a proactive energy service and request models to automatically generate services and  requests according to the spatio-temporal and energy usage patterns of the providers and consumers respectively. The proactive composition approach anticipates the requirements of energy consumers and plans when, where, and how much energy to request for each energy consumer. Preliminary experiments show that proactively generating requests is able to balance the energy supply-demand in the  crowdsourced energy market.


\bibliographystyle{IEEEtran}
\bibliography{IEEEabrv, ref}

\begin{thebibliography}{10}
\providecommand{\url}[1]{#1}
\csname url@samestyle\endcsname
\providecommand{\newblock}{\relax}
\providecommand{\bibinfo}[2]{#2}
\providecommand{\BIBentrySTDinterwordspacing}{\spaceskip=0pt\relax}
\providecommand{\BIBentryALTinterwordstretchfactor}{4}
\providecommand{\BIBentryALTinterwordspacing}{\spaceskip=\fontdimen2\font plus
\BIBentryALTinterwordstretchfactor\fontdimen3\font minus
  \fontdimen4\font\relax}
\providecommand{\BIBforeignlanguage}[2]{{%
\expandafter\ifx\csname l@#1\endcsname\relax
\typeout{** WARNING: IEEEtran.bst: No hyphenation pattern has been}%
\typeout{** loaded for the language `#1'. Using the pattern for}%
\typeout{** the default language instead.}%
\else
\language=\csname l@#1\endcsname
\fi
#2}}
\providecommand{\BIBdecl}{\relax}
\BIBdecl
\renewcommand{\BIBentryALTinterwordstretchfactor}{4}

\bibitem{atzori2010internet}
L.~Atzori \emph{et~al.}, ``The internet of things: A survey,'' \emph{Computer
  networks}, vol.~54, no.~15, pp. 2787--2805, 2010.

\bibitem{ahabak2015femto}
K.~Habak \emph{et~al.}, ``Femto clouds: Leveraging mobile devices to provide
  cloud service at the edge,'' in \emph{IEEE CLOUD}, 2015.

\bibitem{raptis2019online}
T.P. Raptis, ``Online social network information can influence wireless crowd
  charging,'' in \emph{IEEE DCOSS}, 2019.

\bibitem{Previouswork11}
A.~Lakhdari \emph{et~al.}, ``Crowdsourcing energy as a service,'' in
  \emph{Springer ICSOC}, 2018.

\bibitem{lakhdari2020composing}
A.~Lakhdari \emph{et~al.}, ``Composing energy services in a crowdsourced iot
  environment,'' \emph{IEEE TSC}, 2020.

\bibitem{choi2017wearable}
Y.M. Choi \emph{et~al.}, ``Wearable biomechanical energy harvesting
  technologies,'' \emph{MDPI Energies}, 2017.

\bibitem{gorlatova2014movers}
M.~Gorlatova \emph{et~al.}, ``Movers and shakers: Kinetic energy harvesting for
  the internet of things,'' in \emph{ACM SIGMETRICS Performance Evaluation
  Review}, 2014.

\bibitem{lakhdari2020Vision}
A.~Lakhdari \emph{et~al.}, ``Crowdsharing wireless energy services,'' in
  \emph{2020 IEEE CIC}.\hskip 1em plus 0.5em minus 0.4em\relax USA: IEEE, 2020,
  pp. 18--24.

\bibitem{lakhdari2020fluid}
A.~Lakhdari and A.~Bouguettaya, ``Fluid composition of intermittent iot energy
  services,'' in \emph{SCC}.\hskip 1em plus 0.5em minus 0.4em\relax IEEE, 2020.

\bibitem{lakhdari2020Elastic}
A.~Lakhdari \emph{et~al.}, ``Elastic composition of crowdsourced iot energy
  services,'' in \emph{Mobiquitous}.\hskip 1em plus 0.5em minus 0.4em\relax
  EAI, 2020.

\bibitem{abusafia2020reliability}
A.~Abusafia and A.~Bouguettaya, ``Reliability model for incentive-driven iot
  energy services,'' in \emph{Mobiquitous}, 2020.

\bibitem{chaki2020conflict}
D.~Chaki \emph{et~al.}, ``A conflict detection framework for iot services in
  multi-resident smart homes,'' in \emph{IEEE ICWS}, 2020.

\bibitem{gonzalez2008understanding}
M.C. Gonzalez \emph{et~al.}, ``Understanding individual human mobility
  patterns,'' \emph{nature}, 2008.

\bibitem{yang2015mobility}
Z.~Yang \emph{et~al.}, ``Mobility increases localizability: A survey on
  wireless indoor localization using inertial sensors,'' \emph{ACM Computing
  Surveys}, 2015.

\bibitem{abusafia2020incentive}
A.~Abusafia \emph{et~al.}, ``Incentive-based selection and composition of iot
  energy services,'' in \emph{SCC (accepted)}.\hskip 1em plus 0.5em minus
  0.4em\relax IEEE, 2020.

\bibitem{peltonen2015energy}
E.~Peltonen \emph{et~al.}, ``Energy modeling of system settings: A crowdsourced
  approach,'' in \emph{PerCom}.\hskip 1em plus 0.5em minus 0.4em\relax IEEE,
  2015.

\bibitem{deng2016constraints}
S.~Deng \emph{et~al.}, ``Constraints-driven service composition in mobile cloud
  computing,'' in \emph{IEEE ICWS}, 2016.

\bibitem{do2013places}
T.M.T. Do and D.~Gatica-Perez, ``The places of our lives: Visiting patterns and
  automatic labeling from longitudinal smartphone data,'' \emph{IEEE Mobile
  Computing}, 2013.

\end{thebibliography}

\end{document}